\documentclass[conference]{IEEEtran}
\IEEEoverridecommandlockouts
\usepackage{cite}
\usepackage{amsmath,amssymb,amsfonts}
\usepackage{algorithmic}
\usepackage{graphicx}
\usepackage{textcomp}
\usepackage{xcolor}
\usepackage{comment}

\def\BibTeX{{\rm B\kern-.05em{\sc i\kern-.025em b}\kern-.08em
    T\kern-.1667em\lower.7ex\hbox{E}\kern-.125emX}}
    
\begin{document}

\title{Edge Deep Learning Enabled Freezing of Gait Detection in Parkinson's Patients}
%\title{Wireless Sensors with Edge Deep Learning for Detecting and Alerting the Freezing of Gait Symptoms in Parkinson's Patients}

% Authors are commented out as required by double blind review
\author{\IEEEauthorblockN
        {Ourong~Lin\IEEEauthorrefmark{1},
        Tian~Yu\IEEEauthorrefmark{1},
        Yuhan~Hou\IEEEauthorrefmark{1},
        Yi~Zhu\IEEEauthorrefmark{1},
        and~Xilin~Liu\IEEEauthorrefmark{1}\IEEEauthorrefmark{2}
        }
\IEEEauthorblockA{\IEEEauthorrefmark{1}Department of Electrical and Computer Engineering (ECE), University of Toronto, Toronto, ON, Canada \\ \IEEEauthorrefmark{2}Toronto Rehabilitation Institute, University Health Network (UHN), Toronto, ON, Canada \\Email: xilinliu@ece.utoronto.ca}}

\maketitle

\begin{abstract}
This paper presents the design of a wireless sensor network for detecting and alerting the freezing of gait (FoG) symptoms in patients with Parkinson's disease. A novel button pin type sensor node design is developed for easy attachment. Three sensor nodes, each integrating a 3-axis accelerometer, can be placed on a patient at ankle, thigh, and truck. Each sensor node can independently detect FoG using an on-device deep learning (DL) model, featuring a squeeze and excitation convolutional neural network (CNN). The DL model outputs from the three sensor nodes are processed in a central node using a majority voting algorithm. In a validation using a public dataset, the prototype developed achieved a FoG detection sensitivity of 88.8\% and an F1 score of 85.34\%, using less than 20 k trainable parameters per sensor node. Once FoG is detected, an auditory signal will be generated to alert users, and the alarm signal will also be sent to mobile phones for further actions if needed. The sensor node can be easily recharged wirelessly by inductive coupling. The system is self-contained and processes all user data locally without streaming data to external devices or the cloud, thus eliminating the cybersecurity risks and power penalty associated with the wireless data transmission. The developed methodology can be used in a wide range of applications. 

\end{abstract}

\section{Introduction}

Parkinson's disease (PD) is a neurodegenerative disorder that affects more than 8.5 million people worldwide \cite{WHO2019}. Patients with PD experience a multitude of movement disorders. A prevalent disorder in the late stages of PD is freezing of gait (FoG), which impedes a patient's walking and turning, increasing instability and risk of falls and injuries \cite{bikias2021deepfog}. Although there is no known cure for FoG, there are treatment methods, including pharmaceutical treatment and invasive or non-invasive stimulation \cite{armstrong2020diagnosis}. Invasive approaches can achieve high clinical efficacy, but come with risks and adverse effects \cite{bratsos2018efficacy}. Auditory stimulation is a non-invasive approach that is safe to implement and has a high success rate for certain groups of patients \cite{pereira2019music}. However, this approach requires that patients with PD are assisted by clinicians. 
\begin{comment}
\begin{figure}[ht]
\centering
\includegraphics[width=1\linewidth]{./fig/sys_diagram_v5}
\caption{(a) Illustration of the smart sensor network consisting of three distributed sensor nodes with edge intelligence and a central node for detecting and alerting FoG in Parkinson's patients. (b) The high-level block diagram of the system. The sensor nodes and the central node use the same MCU. The accelerometer is installed on the sensor node only and the speaker is installed on the central node only.}
\label{fig:intro}
\end{figure}
\end{comment}

\begin{figure*}[!ht]
    \centering
    \includegraphics[width=.9\textwidth]{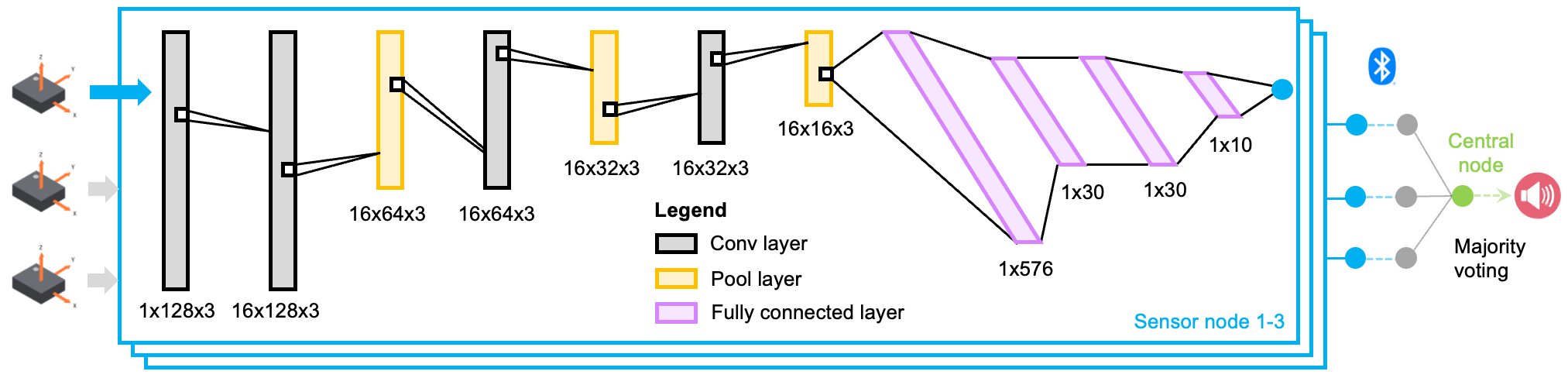}
    \caption{The DL model architecture and the three-sensor majority voting mechanism. Each of the 3 sensor nodes process an 128 x 3 input tensor through a CNN with identical architecture but different weights, and  produce an output ranging in [0,1]. FoG alert simulations are only activated when at least 2 sensor output are greater than 0.4, a low-pass filter that yield the largest area under ROC curve}
    \label{fig:CNN_Arch}
\end{figure*}

Recently, machine learning methods, including deep learning (DL) models, have been developed to detect FoG automatically without involving human in the loop \cite{AlexNet:2020,armstrong2020diagnosis,CNNMLP:2019,CNN:2021}. These methods potentially permit low-cost treatment at home or long-term operation using wearable devices. However, the DL models in existing work demand high computational power, and thus are not suitable for low-power devices. Although processing can be offloaded to the cloud \cite{liu2021edge}, these approaches have major drawbacks in: (1) dependence on the Wi-Fi or cellular network, which prevents offline use, (2) continuous data transmission poses a power penalty, and (3) wireless data transmission poses cybersecurity concerns \cite{liu2021energy}. There is a compelling need for self-contained sensors that can detect FoG locally and generate alerts in real time \cite{liu201512}.

In this work, we fill this important research gap by developing a wireless sensor system that can detect FoG with an edge DL model. The system consists of a central node and several sensor nodes. Each sensor node integrates a low-power microcontroller (MCU) with a wireless module that supports Bluetooth and a 3-axis accelerometer \cite{liu2020fully}. A lightweight DL model with less than 20 k trainable parameters was integrated in the MCU to detect FoG. The sensor node will notify the central node when FoG is detected; the central node will process inputs from all sensor nodes and generate an auditory stimulus via an integrated speaker once pre-defined conditions are met. All nodes are battery powered and can be easily recharged wirelessly by inductive coupling. 

The rest of the paper is organized as follows. Section II first introduces the FoG detection algorithm, presents the wireless sensor hardware design, and discusses the deployment of the algorithm into the hardware. The experimental results are given in Section III and compared with the state-of-the-art work. Section IV concludes the paper.

\section{Methods}

\subsection{Development of the Deep Learning Model}

The DL model was trained and validated using a public dataset reported by M. Bachlin and colleague \cite{CNN:2021}, referred to as the Daphnet dataset in this article. The Daphnet dataset consists of acceleration measurements taken from 10 patients with PD tracked in a controlled environment performing three types of tasks: straight walking, walking with numerous turns, and simulated activity of daily living (ADL) such as fetching coffee and opening doors. Measurements were taken from the ankle, thigh, and truck of patients and sampled at 64 Hz. The measurements were labeled by experts with the FoG status: label 2 is set for freezing, label 1 is for non-freezing, and label 0 is for experiment-irrelevant activities, such as debriefing. Data from the 5th and the 10th patients were omitted from the experiments because they did not experience freezing during the experiments.

%The construction of the model and the processing of the training data were inspired by the work by Mekrksavanich et al., which provided a framework for strong accuracy and precision results while retaining a low model parameter count with a squeeze and excitation convolutional neural network (CNN). These properties prove to be necessary for implementation in embedded systems \cite{CNN:2021}. The precise data processing methods and model architecture are described in the following section.

%The model training and validation used the Daphnet Freezing of Gait dataset \cite{Daphnet:2010}. 

%This dataset presents FoG events with subject acceleration measurements taken from accelerometers on the ankle, thigh, and truck locations. This dataset contains data from 10  \cite{Daphnet:2010}. Each dataset file is indexed by a timestamp distanced by 16ms and consists of the 9 dimensional accelerometer measurement values aforementioned, and finally the encoded label that specifies the status of FoG. 

We adopted a k-fold cross-validation strategy for training the model, with 20\% of the data retained for testing. A hard saturation limit of 5 g was applied to all data, eliminating outliers that could alter the data scaling. The data was filtered through a low-pass filter with a 20 Hz cutoff frequency to reduce noise. This cutoff was chosen because FOG events are best predicted by signals originating from the 0-3 Hz "locomotor" band and the 3-8 Hz "freeze" band \cite{Moore:2013}. Then, the filtered data was normalized to facilitate model training. Finally, the data was segmented into windows of 128 samples with 64 overlap samples between windows, translating to 2-sec windows with a 1-sec overlap. If a window contained one or more irrelevant data points (labeled 0), the window was discarded. Then, windows consisting of more than 40\% freezing points were labeled as freezing, and the remaining windows were labeled as non-freezing. The 40\% threshold was tuned as a hyper-parameter. The labeled windows were then shuffled and used in the training and validating of the model.

We developed a squeeze and excitation CNN model, as shown in Fig. \ref{fig:CNN_Arch}. A three-layer CNN, with 1-D max pooling between convolutional layers, was used to learn feature data from the dataset while reducing the required number of training parameters. 
%This resulted in a total parameter count of 118,459. 
The model was simplified using only native Keras layers to facilitate better translation to a Tensorflow Lite compatible model for implementation on the embedded hardware. This CNN was built and trained using the Tensorflow Keras 2.10.0 library. The output of the convolutional layers was fed into a pair of dense layers separated by a dropout layer and then a final output layer using sigmoid activation for binary classification. Dropout was implemented to increase the stochasticity of training and combat overfitting due to the limited data provided. It was noted that the number of freezing and non-freezing frames in the dataset were not equal. This was expected as the majority of subject time was spent in a non-freezing state. Thus, class weighting was implemented as suggested in the Keras training documentation \cite{Tensorflow:Imbalance:2022}. Additionally, a bias was initialized on the prediction layer to further account for this data imbalance and reduce the required number of training epochs to minimize loss.

\subsection{Wireless Sensor Hardware Design}
All nodes use a 32-bit MCU (nRF52840, Nordic Semiconductor) featuring an ARM Cortex M4 CPU with floating point unit (FPU) running at 64 MHz \cite{liu2017fully,liu2017wireless}. The MCU also integrates a wireless module that supports Bluetooth 5.3 multiprotocol radio, including mesh networking \cite{liu2015pennbmbi}. The MCU integrates 1 MB Flash memory and 256 kB SRAM. The sensor node integrates a 3-axis accelerometer (LSM9DS1, STMicroelectronics), which has a programmable full-scale acceleration from $\pm$ 2g to $\pm$ 16g. A speaker is integrated into the central node, which can produce a programmable auditory stimulus of up to 80 dB. Wireless inductive power transfer is used to charge the battery. A carrier frequency of 250 Hz is used. The coil has an inductance of 60 $\mu$H. A power management module regulates the charging current. An on-device low drop-out regulator (LDO) is used to power the MCU and the sensors. The debugging and programming interface (DPI) allows us to update the program and DL model. 

%Fig. \ref{fig:sensor} shows the photos of an assembled device, which has a diameter of 17 mm. The device has a tie-tack pin on the back for easy attachment. An inductive coil is installed on top of the modules for wireless charging. 
\begin{comment}
\begin{figure}[!ht]
\centering
\includegraphics[width=.8\linewidth]{./fig/pcb}\vspace{-2mm}
\caption{Photos of the (a) front and (b) back of the button pin-type sensor nodes, which have a diameter of 17 mm, on the side of a Canadian coin. Key blocks are highlighted. Epoxy was applied to the coil and the PCB. A cover can be added to the sensor node for better appearance.}
\label{fig:sensor}
\end{figure}
\end{comment}

\subsection{Deployment of the DL Model on the Hardware}

The selected 32-bit MCU nRF52840 is suitable for the computational demand of this work. 
%Arduino BLE 33 was chosen for testing the embedded hardware. This was due to its native accelerometer, as well as the availability of documentation using this device for the implementation of machine learning algorithms. 
Tensorflow Lite was used to convert the DL model developed in Python to a C++ model that can be executed on the MCU. 
%as it directly supports the conversion of standard TensorFlow models . %The implementation code itself was structured around the Tensorflow Lite example program for magic wand detection, which provided a ready-made framework for reading accelerometer values and using them alongside a TensorFlow Lite machine learning model to predict gestures \cite{Magicwand:2022}.
The converted model occupied 478 kB of memory, which was less than the Flash memory integrated in the MCU. To validate hardware deployment and test the performance of the DL model, testing data from the Daphnet dataset was sent to the MCU from a computer host (rather than directly from the accelerometers). A full buffer of 128 values was sent before each prediction was made. This allowed results to be directly compared with model performance with its non-quantized counterpart. A median filter and a first-order discrete low-pass filter with a 20 Hz cutoff frequency were implemented in the MCU to mirror the pre-processing performed during model training and validation. The data was also min-max normalized. On the MCU, the filtered data was stored in a rolling buffer of 128 data points, which was treated as an input window on which the model would make predictions. Due to the simplicity and compressed size of the CNN model developed, minimal changes were required to implement it on the embedded hardware.

\section{Results}\label{section_results}

\begin{table*}[!ht]
\renewcommand{\arraystretch}{1.4} % Default value: 1
\caption{\label{table:comparison}Model Performance Comparison with the State-of-the-Arts.}
\centering
\begin{tabular}{|l|l|l|l|l|l|l|l|}
\hline
Reference & Year & Model Architecture& Accuracy & Sensitivity & Specificity & F1 score & \# Trainable Parameters\\
\hline
Rodriguez-Martin\cite{SVMwrist:2017} & 2017 & SVM with Wrist Sensor &83.66\% & 88.09\% &  80.09\% & --\% & --\\
\hline
San-Segundo\cite{CNNMLP:2019} & 2019 & CNN+MLP & --\% & 92.3\% &92.8\% &94.8\%& 5,001,273\\
\hline
Tautan \cite{AlexNet:2020} & 2020 &1D CNN& --\% & 83.77\% & 81.78\% & --\% & --\\
\hline
Sigcha \cite{Forest:2020} & 2020 &Random Forest& --\% & 87.8\% &  87.6\% & --\% & 298,500\\
\hline
Mekruksavanich  \cite{CNN:2021} & 2021 & Squeeze and Excite CNN& 95.66\% & 95.66\% & --\% & 95.56\% & 32,450\\
\hline
Mesin \cite{SVM:2022} & 2022 &SVM & 88\% & 85.14\% & 88.38\% & 86.73\%& NA\\
\hline
This work (Python) & 2022 &CNN+Majority Voting& 83.00\% & 85.40\% & 82.70\% & 85.50\%& 19,995 each node\\
\hline
This work (Embedded) & 2022 &CNN+Majority Voting & 81.48\% & 88.80\% & 80.71\% & 85.34\%& 19,995 each node\\
\hline
\end{tabular}
\end{table*}

We used accuracy, sensitivity, specificity, and F1 score as metrics for testing the performance of the developed DL model. These metrics are defined as follows:
\begin{equation}
    Accuracy = \frac{TP + TN}{TP+TN+FP+FN}
\end{equation}

\begin{equation}
    Sensitivity = \frac{TP}{TP+FN}
\end{equation}

\begin{equation}
    Specificity = \frac{TN}{TN + FP}
\end{equation}

\begin{equation}
    F1 = \frac{2TP}{2TP + FN + FP}
\end{equation}
where $TP$ is the true positive assessment, $TN$ is the true negative assessment, $FP$ is the false negative assessment and $FN$ is the false negative assessment. 
%These metrics are commonly used in the evaluation of medical tests where the true positive rate may be significantly lower than the true negative rate, and thus accuracy may not be an accurate measurement. 
Sensitivity and specificity give a sense of the propensity of the model for the prediction of $TP$ and $TN$, respectively. F1 score is measure of a test's accuracy and is preferable when the dataset has imbalanced class distribution, such as in our case.

\begin{figure}[!ht]
    \centering
    \includegraphics[width=.45\textwidth]{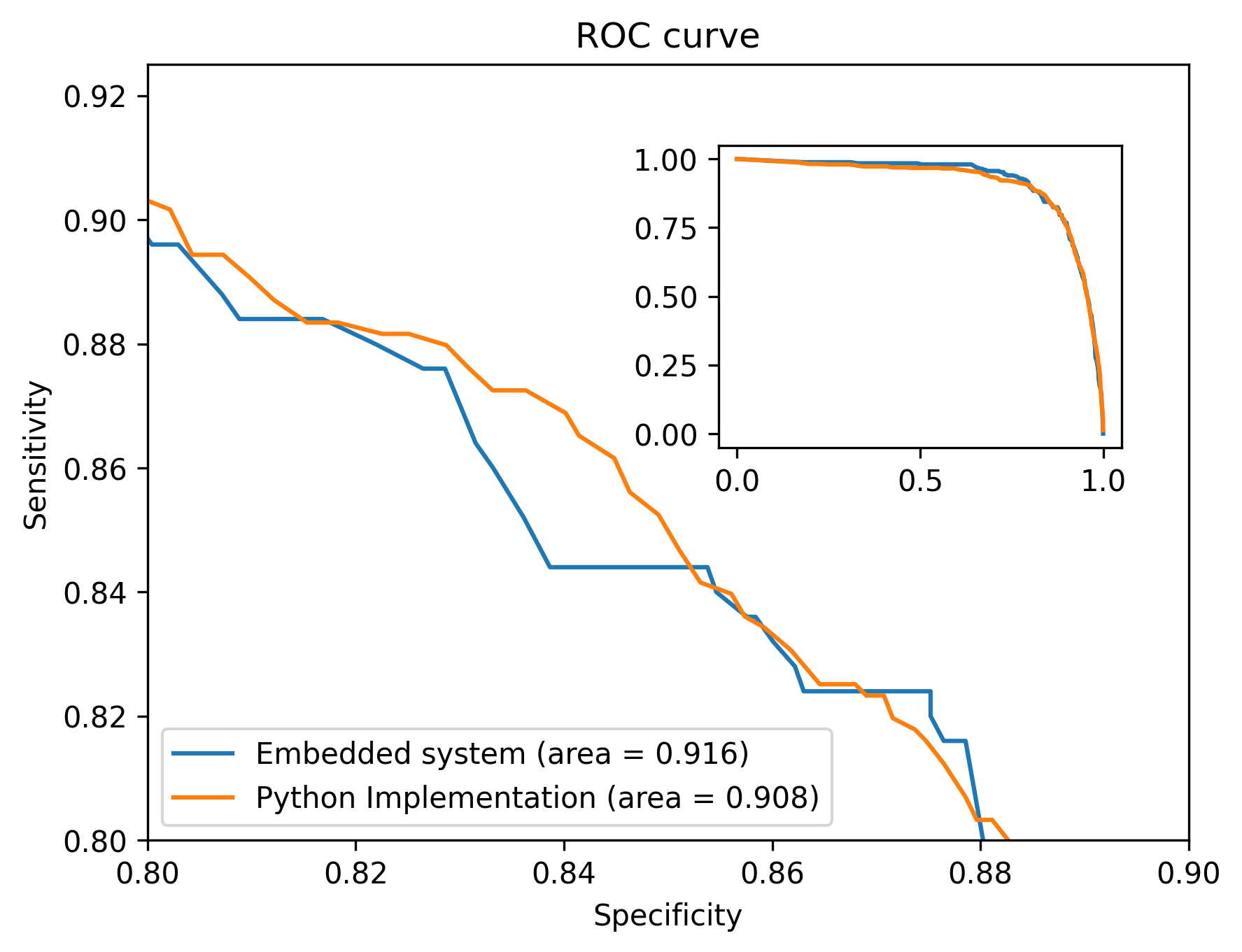}
    \caption{ROC results at sensitivity vs specificity. Since sensitivity is of the main concern, the best performance happens at 88.8\% and 80.71\% for sensitivity and specificity, respectively.}
    \label{fig:ROC}
\end{figure}

These metrics were obtained with a four-fold cross-validation on an aggregated set of patient data from the Daphnet dataset. Table \ref{table:comparison} shows the metrics of the developed model, before and after the quantization and hardware deployment. Fig. \ref{fig:ROC} shows the receiver operating characteristic (ROC) curves of the model before and after hardware implementation. 
Performance metrics after hardware deployment are comparable to the Python implementation, indicating that the model translated successfully into the MCU. The development of the DL model in this work was limited by the hardware resources available in the MCU. As a result, we exclusively used native Keras layers. However, even with hardware constraints, our model was able to achieve a performance that is comparable to the state-of-the-art works. If resources permit, long short-term memory (LSTM) models could also be effective, as they take advantage of the time-series nature of the data \cite{liu2021system}. In addition, we could use data augmentation techniques to further improve model performance as well as generalizability.

The nRF52840 MCU operates with a constant 3.3 V power supply. The measured current consumption during the inference of the DL model was 21 mA and the inference time was 0.21 s. Since the inference frequency of the model is 1 Hz, the active time of the DL engine is 21 \%, resulting in a average current of 4.4 mA. The current consumption of the accelerometer is less than 0.1 mA. The total sensor node device consumes less than 5 mA current, including BLE wireless communication. The wireless charging function of the device has also been fully validated on the bench. The integrated coil can provide a charging current of up to 150 mA with a coupling distance of 1 to 3 cm.

%In summary, this research has shown in preliminary steps that it is possible to implement a FOG detection machine learning algorithm trained in Python on a resource-constrained embedded system without significant loss in performance. Future goals would be to further develop the absolute performance metrics of the detection algorithm to be comparable to current state-of-the-art methods.

\section{Conclusion}
This paper presents the design of a wireless sensor network for detecting and alerting FoG using edge DL. A novel button pin type wireless sensor node is developed. A light weighted DL model was developed and deployed in distributed sensor nodes. The model was validated using a public dataset and achieved a performance comparable to that of state-of-the-art work without hardware implementation. 

In future work, we plan to use the developed wirelss sensor nodes to collect data from healthy subjects and patients with PD, and further optimize the DL model based on the data we collected. In addition, the developed wireless sensors with edge DL can be used in other pre-clinical and clinical experiments, and hold promise in improving the quality of life of a large patient populations with a variety of neurological disorders.

\bibliographystyle{IEEEtran}

\bibliography{ref}

\end{document}